\documentclass[reprint,amsmath,amssymb,pra,longbibliography]{revtex4-1}

\usepackage{graphicx,multirow,color}

\begin{document}

\preprint{APS/123-QED}

\title{Computing resonant modes of circular cylindrical resonators by\\
  vertical mode expansions}

\author{Hualiang Shi}
 \affiliation{School of Science, Hangzhou Dianzi University, Hangzhou,
   Zhejiang, China} 

\author{Ya Yan Lu}
 \email{Corresponding author: mayylu@cityu.edu.hk}
\affiliation{Department of Mathematics, City University of Hong Kong,
  Hong Kong, China} 
 
\date{\today}

\begin{abstract}
Open subwavelength cylindrical resonators of finite
height are widely used in various photonics applications. Circular cylindrical
resonators are particularly important in nanophotonics, since they are 
relatively easy to fabricate and can be designed to exhibit different
resonance effects. In this paper,  an efficient and robust
numerical method is developed for computing resonant modes of circular cylinders which may have a few layers and may be embedded in a
layered background. The resonant modes are complex-frequency 
outgoing solutions of the Maxwell's equations with no sources or
incident waves. The method uses field
expansions in one-dimensional (1D) ``vertical'' modes to reduce the
original three-dimensional eigenvalue problem to 1D problems, and uses
Chebyshev pseudospectral method to compute the 1D modes and set up
the discretized eigenvalue problem. In addition, a new iterative
scheme is developed so that the 1D 
nonlinear eigenvalue problems can be reliably solved. 
For metallic cylinders, the resonant modes are calculated based on 
analytic models for the dielectric functions of metals. 
The method is validated by comparisons  with existing numerical
results, and it is also used to explore subwavelength dielectric
cylinders with high-$Q$ resonances and analyze gold nanocylinders. 
\end{abstract}

\maketitle

\section{Introduction}

Metallic or dielectric circular cylinders of finite height are widely
used as optical resonators in photonics applications \cite{lala18}. Depending on
their material, size and aspect ratio, circular cylinders are used in
integrated photonics as microdisk resonators~\cite{soltani}, in
plasmonics as metallic nanoparticles, and in metasurfaces as building
blocks~\cite{gene17,khor17,su18,shre18}.  Due to their simple geometry, circular
cylinders are relatively easy to fabricate, and they are capable of
creating strong local fields that are useful for lasing, 
sensing, Raman scattering, nonlinear optics, and quantum optics~\cite{lala18}.  To
design resonators of proper material, size and aspect ratio and to
analyze their applications, it is essential to calculate the resonant
modes accurately. A resonant mode (also called resonant state or 
quasinormal mode) is a
complex-frequency solution of the source-free Maxwell's equations
satisfying an outgoing radiation condition.  Some interesting resonant
modes may exist at special geometric parameter values only.  Recently,
it was found that subwavelength dielectric cylinders of particular
aspect ratio can have high-$Q$ resonant modes~\cite{rybin17} and these
modes can be used to enhance second harmonic generation~\cite{carl18}.
To find desired resonant modes for various applications, a robust,
accurate and efficient numerical method is needed.  For metallic
cylinders, the dielectric function depends strongly on the frequency.
Since the resonant frequencies are complex, it is necessary to
extend the dielectric function to the complex frequency plane using
proper analytic models.

For dielectric cylindrical resonators, numerical methods that give the
correct $Q$-factors have appeared since
1980's~\cite{tsuji83,glisson83}.  Currently, the most widely used
method is the finite element method (FEM) with perfectly matched
layers (PMLs)~\cite{hyun97,hwang98,kim09}. FEM is very versatile and its
adaptive version is 
well-suited to analyze structures with complex geometries
\cite{bao05}. PML is a widely used technique for truncating unbound domains in
numerical simulations of waves~\cite{pml94}. For dielectric structures
where the 
material dispersion can be ignored, FEM gives a linear matrix
eigenvalue problem that can be solved using standard numerical linear
algebra techniques. For dispersive media, the eigenvalue problem is
nonlinear, but it can be linearized by using auxiliary functions if an analytic model for the 
dielectric function is available \cite{raman10,yan18}. Typically, FEM
gives rise to large matrices and it is not as efficient as desired.  
More efficient methods can be developed by taking advantage of the
special features of the structure. The boundary integral equation
(BIE) method is suitable for structures with a piecewise constant dielectric
function~\cite{glisson83,powell14}, but it is complicated to implement
when the 
cylinder and/or its surrounding have multiple layers.  The Fourier modal method
(FMM), also called rigorous coupled wave analysis (RCWA),  is widely used in diffraction analysis of layered periodic
structures~\cite{li97,lala01,gran02}, and it has been extended to
computing resonant modes of non-periodic
structures~\cite{lala04}. When
applied to circular cylinders, the standard FMM \cite{lala04,Sauvan:13} uses
vectorial modes that are functions of the two horizontal variables
(perpendicular to the cylinder axis) and avoids a discretization in
the vertical variable $z$ (along the cylinder 
axis). To take advantage of the rotational symmetry of circular
cylinders, two special FMMs have been developed. The method of
Armaroli {\it et al.}~\cite{Armaroli:08}  and Bigourdan {\it et al.}
\cite{Bigourdan:14} uses  one-dimensional (1D) 
modes that depend on $z$ and analytic 
solutions in the horizontal radial variable $r$ and azimuthal angle
$\theta$. The method of Li {\it et al.}~\cite{Li:14} uses 1D modes
that depend on $r$ and analytic solutions in $z$ and $\theta$.  
All versions of BIE and FMM give rise to fully nonlinear eigenvalue problems.   

In this paper, we develop a simple one-dimensional (1D) mode expansion
method for analyzing circular cylindrical resonators. Similar to the
method of Armaroli {\it et al.} \cite{Armaroli:08}, we use 1D modes that
depend on $z$ and analytic solutions in $r$ and $\theta$. 
Instead of Fourier series, we use the Chebyshev pseudospectral
method~\cite{tref} to discretize $z$, calculate the 1D modes, and set up the nonlinear matrix
eigenvalue problem. Our choice is motivated by
the advantage of the Chebyshev pseudospectral method
shown in numerical studies of diffraction gratings \cite{dawei11,granet12}. 
Our method is applicable to multilayered
cylinders embedded in a multilayered surrounding medium. It also
gives rise to nonlinear matrix eigenvalue problems, but the 
matrix size is small. In addition, we develop a robust procedure to
reduce the nonlinear matrix eigenvalue problem to a scalar equation,
so that the complex frequencies of the resonant modes are simply
solutions of the scalar equation. For metallic cylinders, analytic
models for the dielectric functions of metals are needed. For gold, we
show that the critical point (CP) model~\cite{Etchegoin:06,Erratum:07}
gives satisfactory results. Numerical examples are presented to
validate and illustrate our method.

\section{Vertical mode expansions}

We consider a circular cylinder of radius $a$ and height $h$ 
with  its bottom in the $xy$ plane (at $z=0$) and its axis
aligned with the $z$ axis. The dielectric function in the cylindrical
region given by $r < a$ ($r$ is the horizontal
radial variable) is allowed to be a general function of $z$ and
$\omega$, i.e.,  $\varepsilon = \varepsilon^{(0)}(z, \omega)$, where 
$\omega$ is the angular frequency. The medium outside
the cylindrical region can also be layered and its dielectric function
is given by $\varepsilon = \varepsilon^{(1)}(z, \omega)$ for $r > a$. 
In addition, we assume both $\varepsilon^{(0)}$ and $\varepsilon^{(1)}$ become the
same constants for $z >  h$ and for 
$z <  0$, respectively. 

For scattering problems with a given incident wave at a given real
frequency $\omega$, the vertical mode expansion method (VMEM) is very
natural and easy to implement~\cite{Xun:15}. After expanding the
incident wave to components that depend 
on the horizontal angle $\theta$ as $e^{ i m \theta}$ for integers $m$, the
original three-dimensional (3D) problem is reduced to independent 
two-dimensional (2D) problems in $r$ and $z$. For each
$m$, the wave field inside and outside the cylindrical region
can be further expanded in corresponding vertical modes
which are functions of $z$. The expansion coefficients satisfy a
linear system with a $(4N)\times (4N)$ coefficient matrix, where $N$
is the number of points for discretizing $z$. Different approaches
can be used to solve the vertical modes and to set up the linear
systems. The VMEM of \cite{Xun:15} is based on the Chebyshev
pseudospectral method \cite{tref}.

We use VMEM to formulate a nonlinear eigenvalue problem for
resonant modes.  
With a discretization in $z$, the 1D structure given by
$\varepsilon^{(l)}$ (for $l=0$ or $1$), has $2N$
numerically calculated vertical modes  
$\phi_j^{(l,p)}(z)$ with propagation constants $\eta_j^{(l,p)}$ for $j
\in \{1, 2, ..., N\}$ and $p \in \{ e, h \}$. The cases $p=e$
and $p=h$  correspond to the $E$ and $H$ polarizations, respectively.
These vertical modes depend on $\omega$. If a resonant mode depends on
$\theta$ as $e^{ i m \theta}$, its 
vertical components can be approximated by
\begin{eqnarray*}
  H_z = e^{i m \theta} \sum_{j=1}^N c_{j,m}^{(0,e)} \phi_j^{(0,e)}(z)
\frac{ J_m (\eta_j^{(0,e)} r)}{   J_m (\eta_j^{(0,e)} a)},   \  r < a,  \\
  E_z = \frac{e^{ i m \theta} }{\varepsilon^{(0)}(z)} \sum_{j=1}^N 
   c_{j,m}^{(0,h)} \phi_j^{(0,h)}(z) \frac{ J_m (\eta_j^{(0,h)} r)}{   J_m 
   (\eta_j^{(0,h)} a)}, \ r < a, \\
 H_z = e^{i m \theta} \sum_{j=1}^N c_{j,m}^{(1,e)} \phi_j^{(1,e)}(z)
\frac{ H^{(1)}_m (\eta_j^{(1,e)} r)}{H^{(1)}_m (\eta_j^{(1,e)} a)},   \  r > a,  \\
  E_z = \frac{e^{ i m \theta} }{\varepsilon^{(1)}(z)} \sum_{j=1}^N 
   c_{j,m}^{(1,h)} \phi_j^{(1,h)}(z) \frac{H^{(1)}_m (\eta_j^{(1,h)}
   r)}{H^{(1)}_m(\eta_j^{(1,h)} a)}, \ r > a, 
\end{eqnarray*}
where $J_m$ is the Bessel function of first kind and order $m$, $H_m^{(1)}$ is the Hankel
function of first kind and order $m$. The horizontal components
$H_\tau$ and $E_\tau$ (tangential to the boundary of the cylinder at $r=a$) can also be written
down, and they involve the derivatives of $\phi_j^{(l,p)}(z)$
\cite{Xun:15}. The continuity of $H_z$, $E_z$, $H_\tau$ and $E_\tau$
at $r=a$ and the $N$ discretization points of $z$ gives rise to a homogeneous
linear system 
\begin{equation}
  \label{linsys}
{\bf A}_m (\omega)\,  {\bf c}_m = {\bf 0},
\end{equation}
where ${\bf c}_m$ is a column vector of length $4N$ for $c_{j,m}^{(l,p)}$,
$j \in \{1, 2, ..., N\}$, $l \in \{ 0, 1\}$ and $p \in \{ e, h\}$. 
Since all $\phi_j^{(l,p)}$ and $\eta_j^{(l,p)}$ depend on $\omega$, 
the matrix ${\bf A}_m$ also depends on
$\omega$. Equation~(\ref{linsys}) is a fully nonlinear matrix
eigenvalue problem. A resonant mode
corresponds to a complex $\omega$ such that ${\bf A}_m$ is singular.
The wave field of the mode can be constructed from a non-zero vector
${\bf c}_m$ satisfying Eq.~(\ref{linsys}). 

Notice that the right hand side of Eq.~\eqref{linsys} is zero, since 
resonant modes are nonzero solutions without incident waves and
sources. For scattering problems with a given
incident field at a given frequency, the VMEM~\cite{Xun:15} gives rise to 
\begin{equation}
  \label{linsys2}
{\bf A}_m (\omega)\,  {\bf c}_m = {\bf b}_m, 
\end{equation}
where ${\bf b}_m$ is a vector with four blocks related to the 
$z$ and $\tau$ components of electromagnetic fields of some reference 
solutions (induced by the incident wave), and each block is a vector of 
length $N$ corresponding to the $N$ discretization points of $z$.

Nonlinear eigenvalue problems can be solved by local iterative methods
or global contour integration methods~\cite{asak09,beyn12}. A local 
iterative method relies on a scalar function $f(\omega)$, such that
$f(\omega) = 0$ if and only if ${\bf A}_m(\omega)$ is singular. Choices
of $f$ include the determinant of ${\bf A}_m$, the smallest singular
value of ${\bf A}_m$, the smallest eigenvalue (in magnitude) of ${\bf
  A}_m$, etc. The determinant is usually not a good indicator for
singularities of a matrix, unless the size of the matrix is very 
small. The smallest singular value or eigenvalue are better 
indicators, but they can still be difficult to use if the matrix ${\bf 
  A}_m$ is ill-conditioned (close to singular) even when $\omega$ is 
away from a complex resonant frequency. Cheng {\it et al.}
\cite{cheng04} suggested to use 
\begin{equation}
  \label{ourf}
 f(\omega)= \frac{1}{ {\bf a}^T{ \bf A}_m ^{-1} {\bf b}}
\end{equation}
where ${\bf a}$ and ${\bf b}$ are given vectors independent of
$\omega$. If ${\bf a}$ and ${\bf b}$ are chosen randomly, as suggested
by the authors of \cite{cheng04}, the function $f$ above can be rather
oscillatory, then an iterative method may have difficulty to 
converge, even when a good initial guess is available. 

The contour integration methods are more robust. They can be used to
calculate all resonant modes inside a domain in the complex $\omega$
plane, without 
the need for any initial guesses. Equation~(\ref{ourf}) suggests that a
solution $\omega$ of $f(\omega)=0$ is a pole of a complex
function $g(\omega) = {\bf a}^T{ \bf A}_m ^{-1}
{\bf b}$, assuming ${\bf A}_m$ is analytic in $\omega$ and $g$ is
analytic in $\omega$ except at the poles corresponding to the complex
resonant frequencies. Therefore, contour integrals can be used to
determine the poles of $g$ based on the residue theorem. The contour
integration methods of \cite{asak09,beyn12} are more robust since they
replace the vectors ${\bf a}$ and ${\bf b}$ by
matrices, but they are not very efficient, since they need to evaluate the integral on the chosen
contours to high accuracy and these contours cannot be too close to the
complex resonant frequencies.

We use a local iterative method based on the $f(\omega)$ 
given in Eq.~\eqref{ourf}, but  choose ${\bf a}$ and
${\bf b}$ as simple column vectors with only one or two nonzero
entries. The vectors ${\bf a}$ and ${\bf b}$ are chosen 
 such that $f(\omega)$ is smooth near the complex resonant
frequency. 
 Consider $E_z$ and $H_z$ along the vertical boundary of the
cylinder at $r=a$. If $H_z$  is expected to be strong at $z=z_l$
(one of the discretization points of $z$), we
can put a nonzero entry $1$  in the vector ${\bf b}$ at the position
corresponding to  $H_z$ at $z_l$. If $H_z$ is expected to have a
significant overlap with the first 
$E$-polarized vertical mode, we place a nonzero entry $1$  in the vector
${\bf a}$ to pick up 
the coefficient  of $\phi_1^{(0,e)}$. In that case, ${\bf a}^T {\bf A}_m^{-1}
{\bf b} = c_{1,m}^{(0,e)}$  and $f(\omega)
= 1/c_{1,m}^{(0,e)}$. If $H_z$ has a more significant overlap with the
second vertical mode, we choose ${\bf a}$ such that $f(\omega) =1/
c_{2,m}^{(0,e)}$. Similarly, if $E_z$ is the dominant $z$ component,
a nonzero entry of ${\bf b}$ is put in the block corresponding to
$E_z$,  ${\bf a}$  is chosen such that $ f(\omega) =
1/c_{j,m}^{(0,h)}$ where $j$ is usually 1 or 2, depending on which 
vertical mode has a more significant overlap with $E_z$.  
If the structure has a reflection symmetry in $z$, then the resonant
mode is either symmetric or antisymmetric in $z$, we can use a vector
${\bf b}$ with two (symmetrically positioned) nonzero entries, either $1$ and $1$, or $1$ and
$-1$, to excite symmetric or antisymmetric modes, respectively. 
The equation $f(\omega)=0$ can be solved by standard iterative methods
such as the secant method. With this strategy for choosing ${\bf a}$
and ${\bf b}$, the method exhibits excellent global
convergence,  and resonant modes can be found even when the initial
guesses are not very accurate. 

For resonators with a dispersive material, it is necessary to use an
analytic model for its dielectric function, since a resonant mode has 
a complex frequency, but measured data for the dielectric function are
only available for real frequencies. Analytic models for dielectric
functions of metals and other dispersive materials are widely used in
time-domain numerical simulations. The simplest one is
the Drude model, but it is only accurate in a limited frequency
range. The multi-pole Lorentz-Drude models are more appropriate
\cite{raman10,yan18}. For gold, the CP model is only slightly more
complicated than the Drude model, and it gives a good fit for a wide
range of  frequencies \cite{Etchegoin:06,Erratum:07}.  
Some details on the CP model are given in Appendix. 
Notice that all analytic models are obtained by fitting measured
data for real frequencies, it is not clear how accurate these models
are for complex frequencies. It is possible that fitting
real-frequency data with too many terms can only give less accurate
approximations for complex frequencies. We believe the CP model is 
highly appropriate for computing resonant modes of gold resonators in
the optical frequency range.

\section{Dielectric resonators}

To validate  and illustrate our method, we present a few numerical
examples for cylindrical dielectric resonators in this section. 
The first example is a microdisk with a dielectric constant $\varepsilon=10.24$
surrounded by a dielectric medium with $\varepsilon=2.25$. 
The radius and height of the microdisk are 
$a = 0.77$\,$\mu$m and $h=0.24$\,$\mu$m, respectively. This example was
previously analyzed by Armaroli {\it et al.} \cite{Armaroli:08} and
Li {\it et al.}~\cite{Li:14} using special FMMs with 1D vertical and
radial modes, respectively.  
PMLs are used in these works to periodize the $z$ or $r$ directions.
In our method, the $z$ variable is truncated to an
interval of 
$1.92$\,$\mu$m with a total of five layers. The top and bottom layers
are PMLs with a thickness of $0.6$\,$\mu$m. The middle layer
corresponds to the microdisk of height $h$. Between the PMLs and the
middle layer are dielectric layers of $0.24$\,$\mu$m. 
Since the bottom of the microdisk is in the $z=0$ plane,
  the PML above the microdisk  is a layer from 
  $z_{\rm   pml} = 0.48$\,$\mu$m to $z_{\rm end} = 1.08$\,$\mu$m, where
  $z$ is replaced by
\begin{equation}
  \label{pmlformula}
\hat{z} = z + S \int_{z_{\rm pml}}^z \left( \frac{\tau-z_{\rm 
      pml}}{z_{\rm end} - z_{\rm pml}}\right)^2 d\tau.  
\end{equation}
The PML below the microdisk is similarly defined.
The $z$ variable is discretized by Chebyshev
points in five subintervals with a  total of $N=108$ discretization
points, and the parameter $S$ for the PMLs is
  $S=3+7i$. Our results 
are listed in Table~\ref{tab:die01}
\begin{table}[ht!]
\centering 
\caption{Resonant wavelength $\mbox{Re}(\lambda)$ (in $\mu$m) and 
  quality factor $Q$ for  selected modes of a microdisk resonator.} 
\vspace{1mm}
\begin{tabular}{ |c|c|c|c|c|c|c|}
\hline 
\multirow{2}{*}{Mode} &
 \multicolumn{2}{c|}{Armaroli~\cite{Armaroli:08}} &
           \multicolumn{2}{c|}{Li~\cite{Li:14}} & \multicolumn{2}{c|}{This work} \\ 
\cline{2-7} 
& Re($\lambda$) & $Q$ & Re($\lambda$) & $Q$ & Re($\lambda$) & $Q$ \\
\hline 
$\text{TE}_{1,5}$ & 1.5735 & 16 & 1.5728 & 19 & 1.5729 & 20 \\\hline 
$\text{TE}_{1,6}$ & 1.4019 & 34 & 1.4016 & 41 & 1.4016 & 41 \\\hline 
$\text{TE}_{1,7}$ & 1.2655 & 82 & 1.2665 & 89 & 1.2665 & 90 \\\hline 
$\text{TE}_{1,8}$ & 1.1583 & 175 & 1.1574 & 199 & 1.1574 & 200 \\\hline 
$\text{TE}_{1,9}$ & 1.0694 & 350 & 1.0673 & 457 & 1.0674 & 456 \\\hline 
$\text{TE}_{1,10}$ & 0.9938 & 828 & 0.9914 & 1059 & 0.9915 & 1061 \\
\hline
$\text{TM}_{1,6}$ & 1.3079 & 25 & 1.3052 & 25 & 1.3053 & 25 \\\hline 
$\text{TM}_{1,7}$ & 1.2045 & 52 & 1.1998 & 51 & 1.1998 & 51 \\\hline 
$\text{TM}_{1,8}$ & 1.1122 & 105 & 1.1112 & 107 & 1.1112 & 107 \\\hline 
$\text{TM}_{1,9}$ & 1.0358 & 215 & 1.0356 & 237 & 1.0357 & 238 \\\hline 
$\text{TM}_{1,10}$ & 0.9706 & 536 & 0.9703 & 549 & 0.9704 & 548 \\\hline 
$\text{TM}_{1,11}$ & 0.9132 & 1254 & 0.9130 & 1303 & 0.9131 & 1303 \\
\hline 
\end{tabular}
\label{tab:die01}
\end{table}
for comparison with those of \cite{Armaroli:08} and \cite{Li:14}. For all
three methods, we list the resonant wavelength $\mbox{Re}(\lambda)$,
where $\lambda  =2\pi c/\omega$ is the complex wavelength and $c$ is
the speed of light in vacuum, and the quality factor $Q = -0.5
\mbox{Re}(\omega)/\mbox{Im}(\omega) = 0.5  \mbox{Re}(\lambda)/
\mbox{Im}(\lambda)$.  The quasi-TE  (quasi-TM) 
modes have a dominant $H_z$ ($E_z$) component, and are denoted as 
TE$_{j,m}$ (TM$_{j,m}$), where $m$ is the azimuthal index and $j$ is the
mode index. The case $j=1$ corresponds to a
vertical profile  with a single field maximum located at
the middle of the microdisk. Large values of $m$ correspond to 
whispering-gallery modes with high $Q$-factors. Our results agree very
well with those of Li {\it et al.}~\cite{Li:14}. Notice that the
resonant wavelength decreases as $m$ increases, and only the first mode
in the table, i.e., TE$_{1,5}$,  has a resonant wavelength larger than the
diameter $2a=1.54$\,$\mu$m.

\begin{figure}[htbp]
\centering 
\includegraphics[scale=0.54]{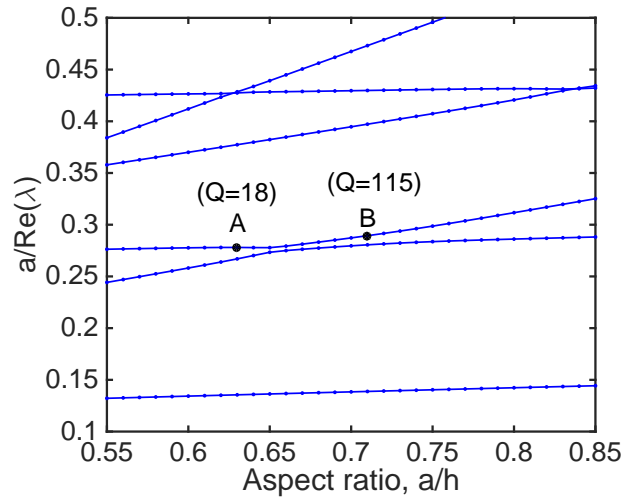}
\caption{Normalized resonant frequencies of a few resonant modes on
  an AlGaAs cylinder of varying aspect ratio.}
\label{fig:highQ}
\end{figure}
\begin{figure}[!htbp]
\centering
\includegraphics[scale=0.35]{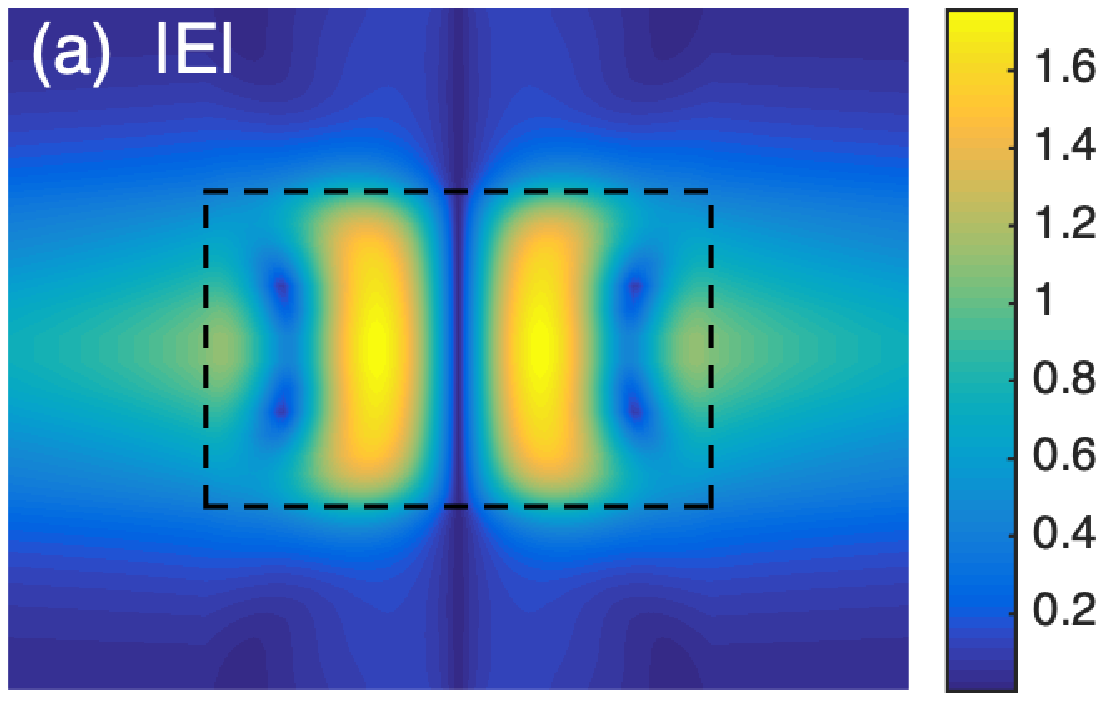} \;
\includegraphics[scale=0.35]{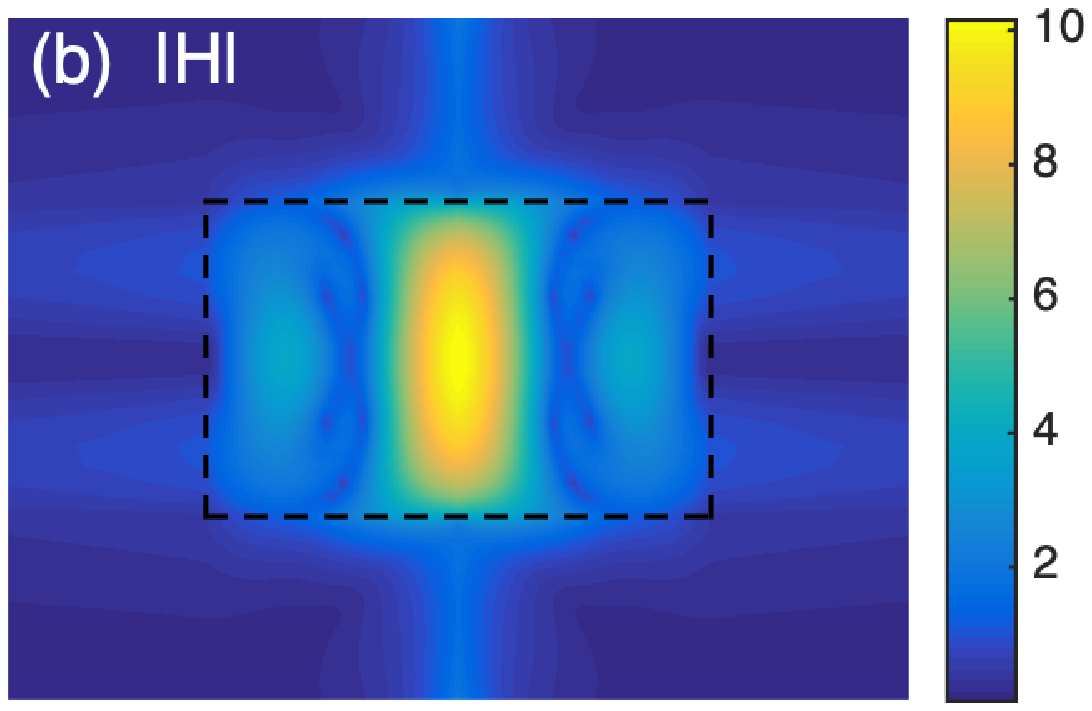} \\
\includegraphics[scale=0.35]{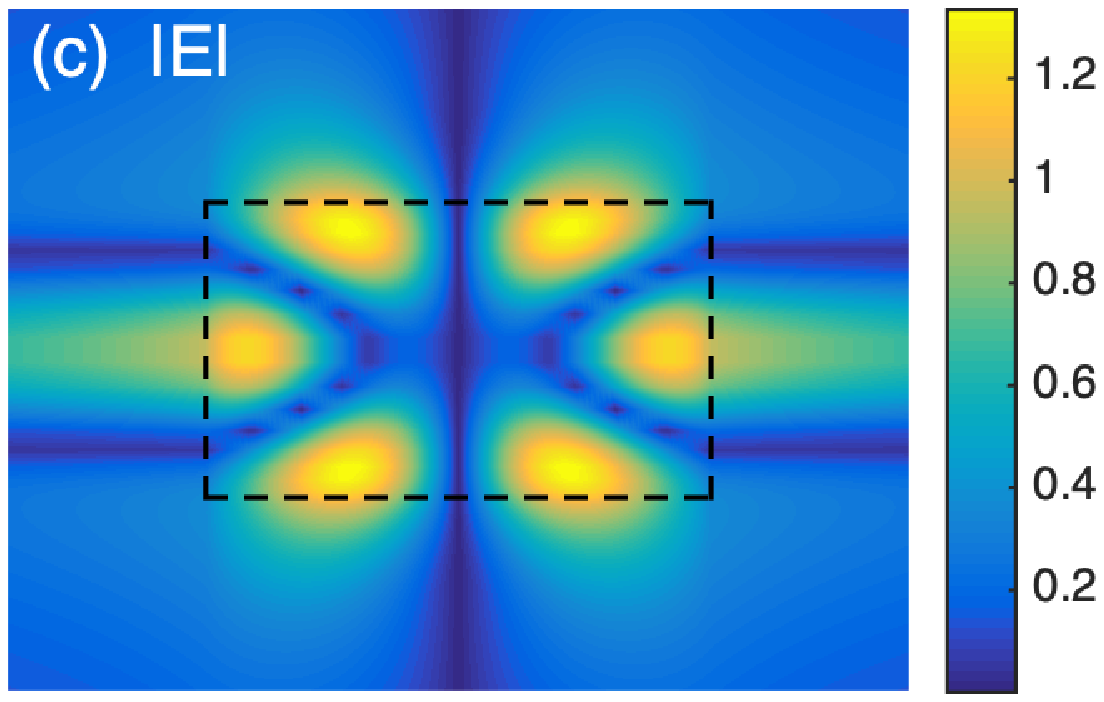} \;
\includegraphics[scale=0.35]{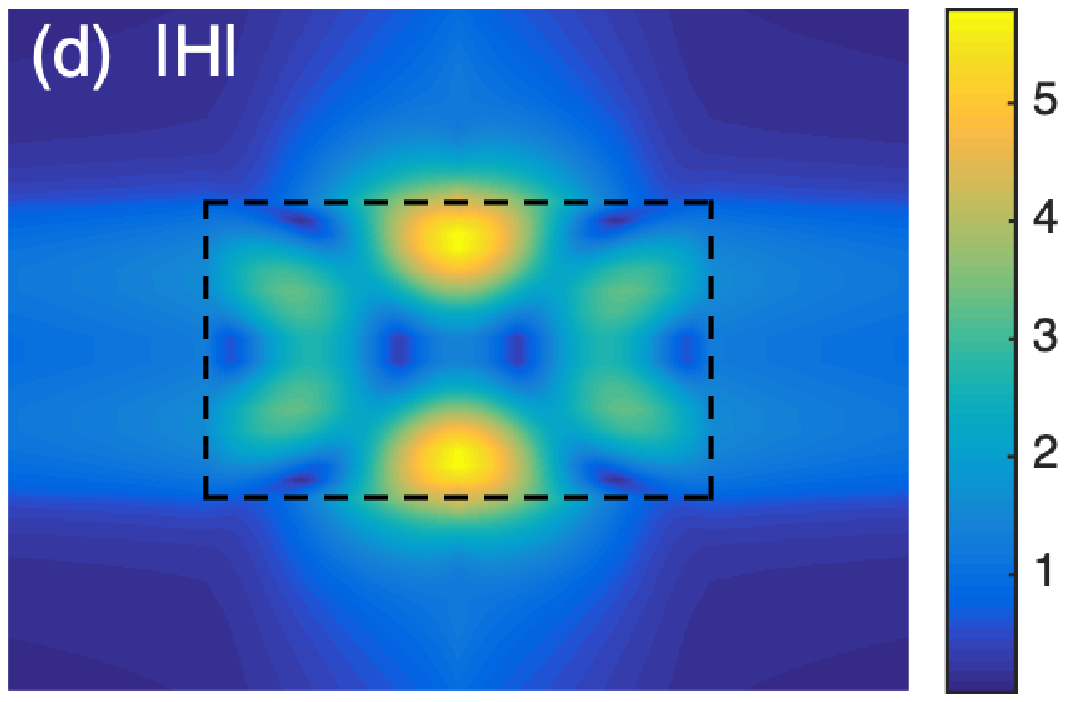}
\caption{Magnitudes of the electric field ${\bf E}$
    and scaled magnetic field ${\bf H}$ on the $xz$ plane for the AlGaAs cylinder
  at points {\sf A} and {\sf B} in Fig. \ref{fig:highQ}, where (a) and (b) are
  for point {\sf A}, (c) and (d) are for point {\sf B}.}
\label{fig:highQ_EMfield}
\end{figure}

Recently, subwavelength dielectric structures supporting high-$Q$
resonances have been designed by relating them to periodic
structures with bound states in the continuum \cite{rybin17}. In
particular, resonant modes  with quality factors over $100$ have been 
found on subwavelength circular cylinders of AlGaAs
and they have been used to enhance nonlinear optical effects
\cite{carl18}. Using our method presented in the previous section,  we
calculate a few  resonant modes for a circular AlGaAs cylinder surrounded by air,
assuming the dielectric constant 
of AlGaAs is  $\varepsilon = 10.73$. In  Fig.~\ref{fig:highQ}, 
we show the first six symmetric quasi-TE modes of
azimuthal order $m=0$ 
for different aspect ratio $a/h$. 
 The vertical axis of Fig.~\ref{fig:highQ}
is the real normalized resonant frequency $\mbox{Re}(\omega)a/(2\pi
c)=  a/\mbox{Re}(\lambda)$. 
Our results agree very well with those of Carletti {\it et al.}
\cite{carl18}.  
In our calculations, the vertical variable $z$ is truncated by PMLs
and discretized by $N=108$ points. In Fig.~\ref{fig:highQ}, two points
are highlighted on  the curve corresponding to the third 
smallest resonant frequency. The points {\sf A} and {\sf B} correspond
to resonant modes with quality factors $Q=18$ and $Q=115$,
respectively. The field profiles for these two points 
are shown in Fig.~\ref{fig:highQ_EMfield},
where ${\bf H}$ is the magnetic field multiplied by
the free space impedance, so that ${\bf H}$ and electric field ${\bf
  E}$ have the same physical units.  For both ${\sf A}$ and ${\sf B}$,
the resonant wavelength is significantly larger than the diameter and
height of the cylinder. 

\begin{figure}[tbh]
\centering 
\includegraphics[scale=0.35]{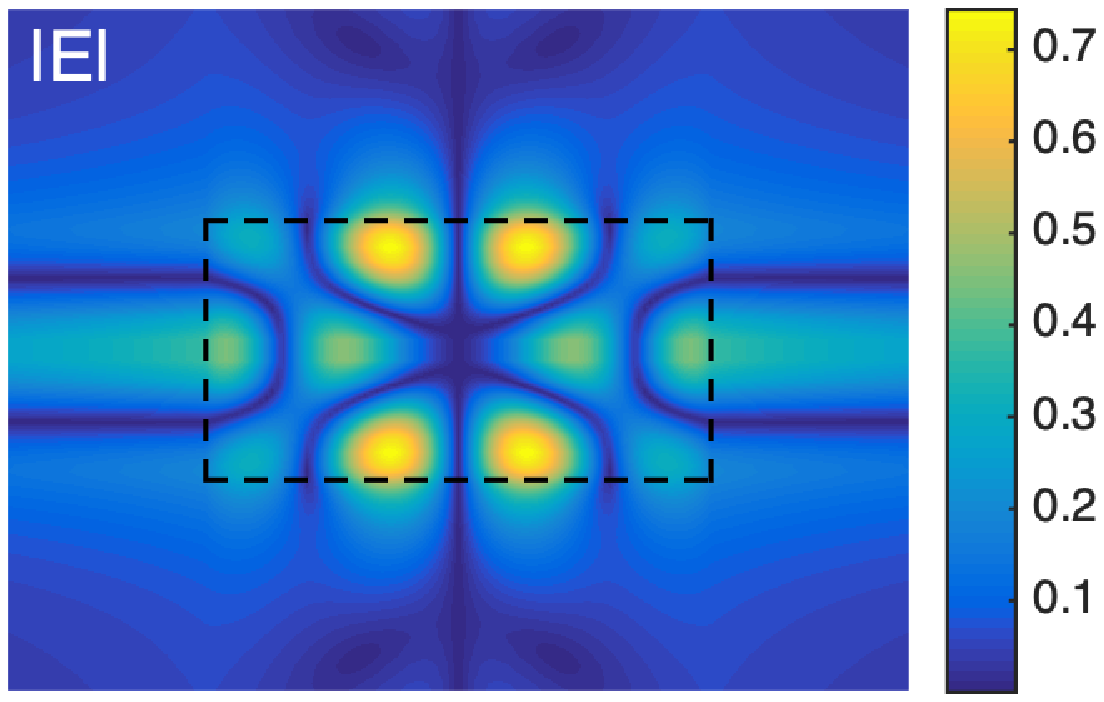} \;
\includegraphics[scale=0.35]{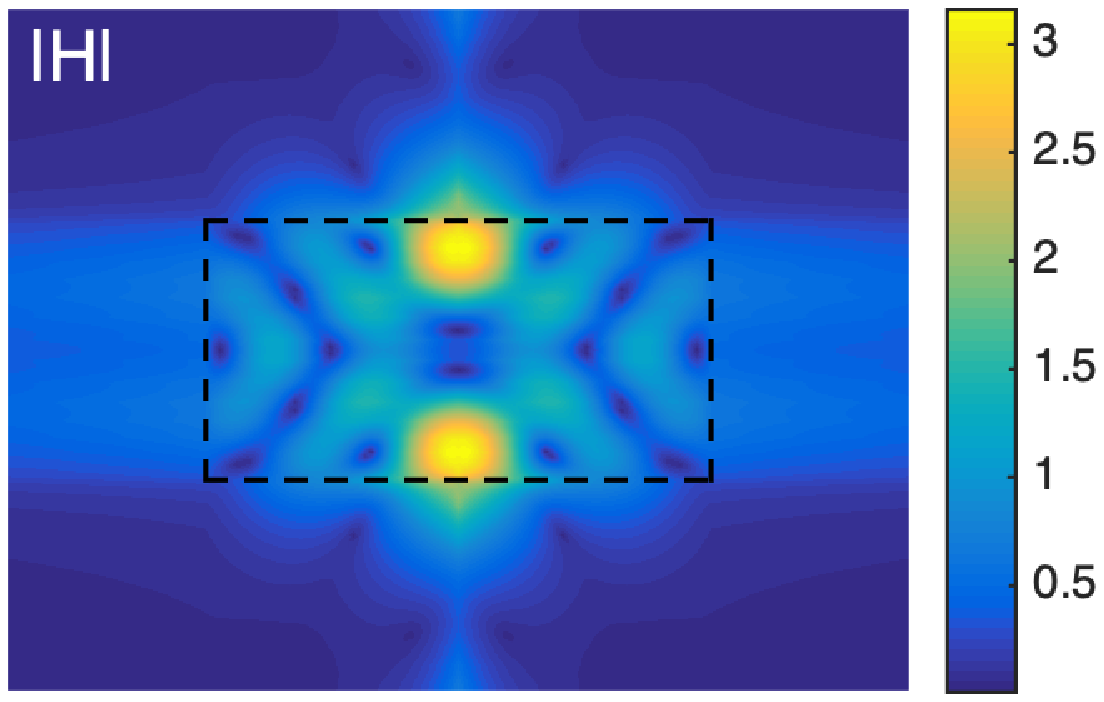}
\caption{Magnitudes of the electric field ${\bf E}$
    (left panel) and
    scaled magnetic field ${\bf H}$ (right panel) on the $xz$ plane of a resonant mode for a
  silicon cylinder with aspect ratio $a/h=0.88211$.}
\label{fig:nomiddle}
\end{figure}

If the dielectric constant of the cylinder is further increased, the quality
factors of the resonant modes can be even larger. For example,
if the dielectric constant of the cylinder is changed to 
$\epsilon = 11.56$ (for silicon) and the surrounding medium is still
air ($\epsilon = 1$), there is a high-$Q$ resonant mode for aspect
ratio $a/h = 0.88211$ 
and the  quality factor  is $Q \approx 179.427$. 
The normalized complex frequency of this resonant mode is 
$\omega a/(2\pi c) = a/\lambda =  0.4256205- 0.001186053i$. 
Its electromagnetic field patterns are shown in Fig.~\ref{fig:nomiddle}. 
Notice that the diameter and height of the cylinder are still smaller than the
resonant wavelength.

\section{Metallic resonators}

In this section, we calculate some resonant modes for circular metallic
cylinders of different sizes. The first example is a gold nanorod
of radius $a=15$\,nm and height $h=100$\,nm, embedded in a dielectric
medium of $\varepsilon=2.25$. This example was previously 
analyzed using a finite element method \cite{Bai:13} and a FMM
\cite{Sauvan:13, Bigourdan:14}.  In these works, a particular resonant mode with azimuthal order 
$m=0$ was carefully studied, while the dielectric function of gold is
approximated by a Drude  model
\[
\varepsilon(\omega) = \varepsilon_{\infty} -
\frac{\omega_p^2}{\omega^2 + i \Gamma \omega},
\]
with parameters $\varepsilon_\infty = 1$, $\omega_p = 1.26 \times
10^{16}$\,rad$/$s  and $\Gamma = 1.41 \times 10^{14}$\,rad$/$s.
These parameters are chosen to fit the measured data of
\cite{Palik:85}. 
The complex wavelength of that mode is
approximately  
$0.9177210+i0.0469092$\,$\mu$m~\cite{Bai:13} or
$0.9173666+i0.0468896$\,$\mu$m~\cite{Bigourdan:14}. 
Using the same Drude model, we calculate the resonant mode with our
method, and obtain the complex wavelength 
$\lambda = 0.9176863 + i0.0469084$\,$\mu$m. 
Our result has an excellent agreement with the FEM
result~\cite{Bai:13} and a good agreement with FMM
result~\cite{Bigourdan:14}. Due to field 
singularity along the sharp edges  and the large field gradient
at the surface of the nanorod, numerical methods typically
exhibit a slow convergence, and it is difficult to assess the accuracy
of these solutions. To obtain our result,  we used $N=265$ points to discretize the $z$ variable
truncated to the interval $(-1, 1.1)$\,$\mu$m, where 
the bottom of the nanorod is at $z=0$. 


Next, we calculate the resonant modes of a gold cylinder with 
radius $a=40$\,nm and height $h=50$\,nm, assuming it is surrounded by
a homogeneous medium with dielectric constant $\varepsilon = 2.25$. In
order to find resonant 
modes with different resonant frequencies, it is desirable to use
an analytic model (for the dielectric function of gold) which is accurate for a
wider frequency range. One possibility is to use the Lorentz-Drude
model \cite{raman10,yan18}. We choose to use the relatively simple CP
model \cite{Etchegoin:06,Erratum:07}. Some details of the CP model are
given in Appendix. It should be pointed out that 
all these models are obtained by fitting measured data for real
frequencies, but what is needed is a formula for the dielectric
function on the complex $\omega$ plane (at least near the real
axis). This is a difficult task, since the measured data on the real
$\omega$ axis have  only limited accuracy,  and more importantly, there is no guarantee that a
formula fitting real $\omega$ data very well remains accurate for complex
$\omega$. From that perspective, a simple formula, such as the CP
model, that fits  the real
$\omega$ data reasonably well over a sufficient large frequency range
is probably the right choice. 
 Based on the CP model, we obtain a symmetric resonant mode of azimuthal order
$m=1$ with complex wavelength $\lambda=0.6369+i0.04402$\,$\mu$m and quality factor $Q=7.2342$.  
For this calculation, the vertical variable $z$ is truncated 
to $(-0.3,0.35)\mu$m by PMLs, and the boundaries between the bottom 
and top PMLs, dielectric layers and the cylinder are located at
$z=-0.1$, $0$, $0.05$ and $0.15$\,$\mu$m. 
The PML above the cylinder is a layer from $z_{\rm
    pml}=0.15$\,$\mu$m to $z_{\rm end} = 0.35$\,$\mu$m, and complex
  variable $\hat{z}$ is defined in Eq.~\eqref{pmlformula} for $S
  =7+5i$. The bottom PML is similar.
The five subintervals of $z$ are 
discretized by $47$, $25$, $13$, $25$ and $47$ points,
respectively. The total number of discretization points for $z$ is 
$N=157$. 

\begin{figure}[htb]
\centering 
\includegraphics[scale=0.5]{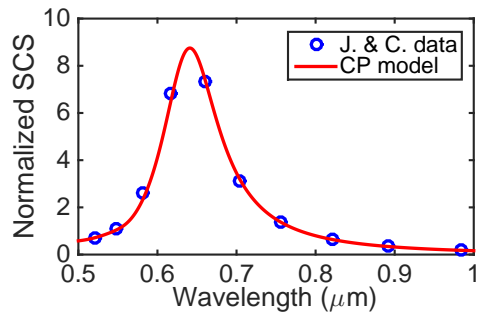}
\caption{Normalized scattering cross section of a gold circular
  cylinder with radius $40$nm and height $50$nm for a normal incident
  plane wave.} 
\label{whatgold}
\end{figure}

In order to provide some justification for our choice of the CP model,
we calculate the scattering spectrum of the gold cylinder for normal
incident plane waves. In Fig.~\ref{whatgold}, 
we show the normalized scattering cross section as a function of the
incident wavelength. The results are obtained using the VMEM for
scattering problems as formulated in \cite{Xun:15}. The red solid line and the blue
circles are results obtained using the CP model
and the measured data of Johnson and Christ \cite{JC:72}. 
Due to the circular geometry of the cylinder, a normal incident plane
wave (with a wavevector parallel to the $z$ axis) produces a scattering
field with an azimuthal dependence of $\sin(\theta)$ and
$\cos(\theta)$. Therefore, the normal incident plane wave 
can only excite resonant modes with azimuthal order $m = \pm
1$.  The peak of the scattering spectrum is located at 
$0.641$\,$\mu$m, and it is close to the resonant wavelength
$\mbox{Re}(\lambda) = 0.6369$\,$\mu$m calculated
earlier. By measuring the difference in wavelengths at which the normalized
scattering cross section reaches its half-maximum, an approximation of
the quality factor can be obtained, and it is about $7.54$. The
agreement with the directly calculated value $Q = 7.2342$ is
acceptable. Since the quality factor is quite small, it is impossible
to accurately extract the resonant wavelength and quality factor from
the scattering spectrum. Based on these calculations, we believe that
the CP model can give satisfactory results for resonant modes of gold
resonators in the optical frequency range. 

\section{Conclusion}
Open circular cylindrical resonators appear in numerous nanophotonics
applications. A special numerical method is developed for computing
resonant modes of (possibly multilayered) circular cylinders
of finite height embedded in a possibly layered background. 
The method relies on expansions of the field in 1D modes which are
functions of $z$, establishes 1D eigenvalue
problems using Chebyshev pseudospectral method, and includes a new
procedure for solving the resulting nonlinear eigenvalue problems. 
 The method is further
applied to determine the aspect ratio of subwavelength silicon
cylinder with a high-$Q$ resonance ($Q\approx 179.427$). It is also used
to analyze a gold nanocylinder. It is shown 
that the resonant wavelength and $Q$ factor calculated directly
using the CP model (for the dielectric function of gold) agree reasonably
well with those extracted from the scattering spectrum. 

Although general numerical methods, such as the FEM, are available for
computing resonant modes even when the media are dispersive, our
method is simple, efficient and robust. For scattering problems, the
VMEM is applicable to more general structures including cylinders with
arbitrary cross sections \cite{hualiang15}, multiple cylinders
\cite{xun16,bowtie} and periodic arrays of cylinders
\cite{hualiang16}.  We are extending the method for computing resonant
modes for such more general structures.

\section*{Acknowledgments}
The first author acknowledges support from the National Natural
Science Foundation of China (Grant No. 11847156). The second author
acknowledges support from the Research Grants  Council of Hong Kong
Special Administrative Region, China (Grant  No. CityU 11304117). 

\section*{Appendix}
\begin{figure}[!htbp]
\centering
\includegraphics[scale=0.3]{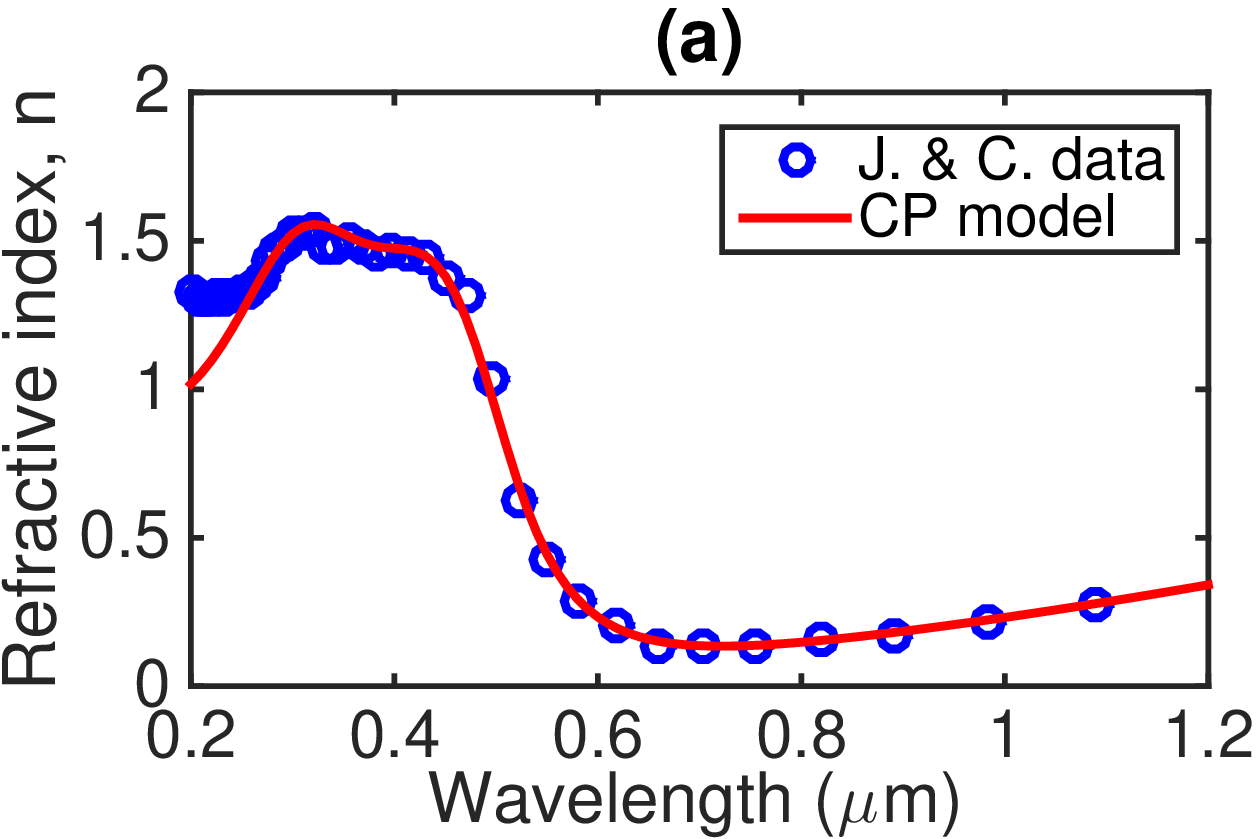}
\includegraphics[scale=0.3]{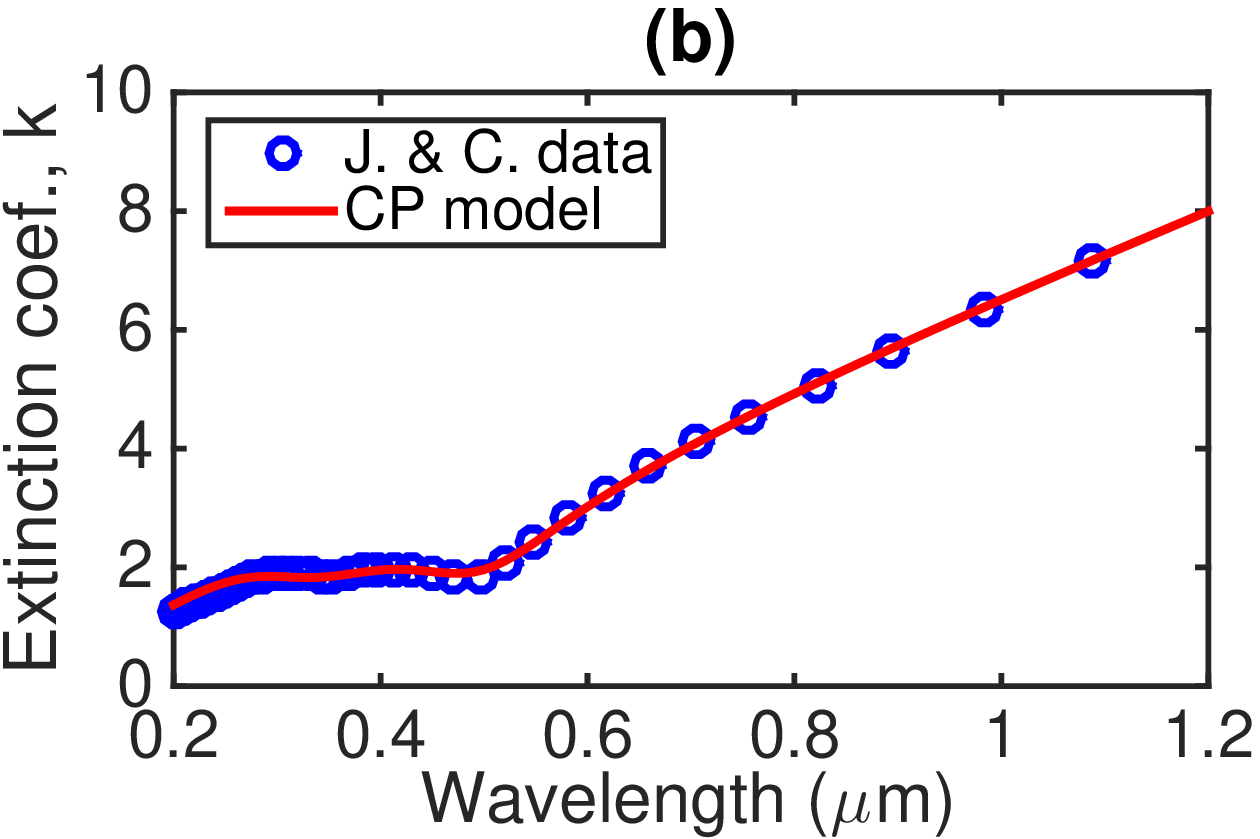} 
\caption{Comparison of the CP model and measured data of 
\cite{JC:72} for gold. Panels (a) and (b) show real and imaginary
parts of  $\sqrt{ \varepsilon} = n+i k$.}
\label{fig:Drude_CP}
\end{figure}

The critical point (CP) model \cite{Etchegoin:06,Erratum:07} for gold  is 
\[
\varepsilon(\omega)  = \varepsilon_{\infty} -
  \frac{\omega_p^2}{\omega^2+i\Gamma\omega} + 
\sum_{j=1}^2 G_j(\omega), 
\]
where the first two terms in the right hand side above is the Drude model, and
\[
 G_j(\omega) =
C_j \left( \frac{e^{i\phi_j}}{\omega_j - \omega - i\Gamma_j}
+ \frac{e^{-i\phi_j}}{\omega_j + \omega + i\Gamma_j}
   \right).
\]
In the above,  $\varepsilon_\infty$, $\omega_p$,  $\Gamma$,
$\omega_j$,  $\Gamma_j$,
$\phi_j$ and $C_j$ are parameters chosen to fit the 
measured data of \cite{JC:72}, and they are 
\begin{eqnarray*}
& \varepsilon_{\infty} = 1.54, & \phi_1 = \phi_2 = -\pi/4, \\
& \omega_p = 1.31815 \times10^{16}, \quad
& \Gamma =  1.29997 \times10^{14}\\ 
& C_1 = 5.09339 \times10^{15}, \quad 
& C_2 = 6.37985 \times10^{15}\\
& \omega_1 = 4.01054 \times10^{15}, \quad 
& \omega_2 = 5.79986 \times10^{15}, \\
& \Gamma_1 = 9.92082 \times10^{14}, \quad 
& \Gamma_2 = 1.77826 \times10^{15}.
\end{eqnarray*}
The unit for $\omega_p$, $\Gamma$, $C_j$, $\omega_j$ and $\Gamma_j$ is
rad$/$s. In Fig.~\ref{fig:Drude_CP}, 
we compare the CP model for gold with the 
data of \cite{JC:72} for the real and imaginary parts of
$\sqrt{\varepsilon} = n+ik$.

\end{document}